\documentclass{article}
\usepackage{ismir,amsmath,cite,url}
\usepackage{graphicx}
\usepackage{color}

\usepackage[defaultsans]{lato}

\usepackage{cite}
\usepackage{amsmath,amssymb,amsfonts}
\usepackage{graphicx}
\usepackage{textcomp}

\usepackage{booktabs}
\usepackage{multirow}
\usepackage{nicefrac}
\usepackage[caption=false]{subfig}
\usepackage[inline]{enumitem}
\usepackage{url}
\usepackage[group-four-digits]{siunitx}
\usepackage[colorinlistoftodos,prependcaption,textsize=tiny]{todonotes}
\usepackage{microtype}

\usetikzlibrary{arrows,automata,shapes}

\graphicspath{{figs/}}

\title{Improved Chord Recognition by Combining Duration and Harmonic Language Models}

\oneauthor
{Filip Korzeniowski and Gerhard Widmer}
{Institute of Computational Perception,\\
Johannes Kepler University, Linz, Austria\\
filip.korzeniowski@jku.at}

\newcommand{\y}{y}
\newcommand{\Yset}{\mathcal{Y}}
\newcommand{\x}{\mathbf{x}}
\newcommand{\s}{s}
\newcommand{\compress}[1]{\mathcal{C}\left(#1\right)}
\DeclareMathOperator*{\argmax}{\arg\!\max}
\newcommand{\seq}[3]{#1_{#2:#3}}
\newcommand{\Y}{\seq{\y}{1}{T}}
\newcommand{\X}{\seq{\x}{1}{T}}

\begin{document}

\maketitle

\begin{abstract}
Chord recognition systems typically comprise an acoustic model that predicts
chords for each audio frame, and a temporal model that casts these predictions
into labelled chord segments. However, temporal models have been shown to only
smooth predictions, without being able to incorporate musical information
about chord progressions. Recent research discovered that it might be the
low hierarchical level such models have been applied to (directly on audio frames)
which prevents learning musical relationships, even for expressive models such
as recurrent neural networks (RNNs). However, if applied on the level of chord
sequences, RNNs indeed can become powerful chord predictors. In this
paper, we disentangle temporal models into a harmonic language model---to be
applied on chord sequences---and a chord duration model that connects the
chord-level predictions of the language model to the frame-level predictions of
the acoustic model. In our experiments, we explore the impact
of each model on the chord recognition score, and show that using
harmonic language and duration models improves the results.
\end{abstract}

\section{Introduction}

Chord recognition methods recognise and transcribe musical chords from
audio recordings. Chords are highly descriptive harmonic features that form
the basis of many kinds of applications: theoretical, such as computational
harmonic analysis of music; practical, such as automatic lead-sheet creation
for musicians\footnote{\url{https://chordify.net/}} or music tutoring
systems\footnote{\url{https://yousician.com}}; and finally, as basis for
higher-level tasks such as cover song identification or key classification.
Chord recognition systems face the two key problems of extracting meaningful
information from noisy audio, and casting this information into sensible
output. These translate to \emph{acoustic modelling} (how to predict a chord
label for each position or frame in the audio), and \emph{temporal modelling}
(how to create meaningful segments of chords from these possibly volatile
frame-wise predictions).

Acoustic models extract frame-wise chord predictions, typically in the form of
a distribution over chord labels. Originally, these models were hand-crafted
and split into feature extraction and pattern matching, where the former
computed some form of pitch-class profiles (e.g.
\cite{muller_making_2009,ueda_hmmbased_2010,mauch_simultaneous_2010}), and the
latter used template matching or Gaussian
mixtures~\cite{cho_improved_2014,fujis hima_realtime_1999} to model these
features. Recently, however, neural networks became predominant for acoustic
modelling~\cite{korzeniowski_feature_2016,humphrey_rethinking_2012,ko
rzeniowski_fully_2016,mcfee_structured_2017}. These models usually compute a
distribution over chord labels directly from spectral representations and thus
fuse both feature extraction and pattern matching. Due
to the discriminative power of deep neural networks, these models achieve superior results.

Temporal models process the predictions of an acoustic model and cast them
into coherent chord segments. Such models are either task-specific, such as
hand-designed Bayesian networks~\cite{mauch_simultaneous_2010}, or general
models learned from data. Here, it is common to use hidden Markov
models~\cite{cho_exploring_2010} (HMMs), conditional random
fields~\cite{korzeniowski_fully_2016} (CRFs), or recurrent neural networks
(RNNs)~\cite{boulanger-lewandowski_audio_2013,sigtia_audio_2015}. However,
existing models have shown only limited capabilities to improve chord
recognition results. First-order models are not capable of learning meaningful
musical relations, and only smooth the
predictions~\cite{cho_relative_2014,chen_chord_2012}. More powerful models,
such as RNNs, do not perform better than their first-order
counterparts~\cite{korzeniowski_futility_2017}. In addition to the fundamental
flaw of first-order models (chord patterns comprise more than two
chords) both approaches are limited by the low hierarchical level they are
applied on: the temporal
model is required to predict the next symbol for each audio frame. This makes
the model focus on short-term smoothing, and neglect longer-term musical
relations between chords, because, most of the time, the chord in the next
audio frame is the same as in the current one. However, exploiting these
longer-term relations is crucial to improve the prediction of chords. RNNs, if
applied on \emph{chord sequences}, are capable of learning these relations, and
become powerful chord predictors~\cite{korzeniowski_largescale_2018}.

Our contributions in this paper are as follows:
\begin{enumerate*}[label=\roman*)]
\item we describe a probabilistic model that allows for the integration of
      chord-level language models with frame-level acoustic models, by connecting
      the two using chord duration models;
\item we develop and apply chord language models and chord duration models
      based on RNNs within this framework; and
\item we explore how these models affect chord recognition results, and show
      that the proposed integrated model out-performs existing temporal models.
\end{enumerate*}

\section{Chord Sequence Modelling}\label{sec:chord_sequence_modelling}

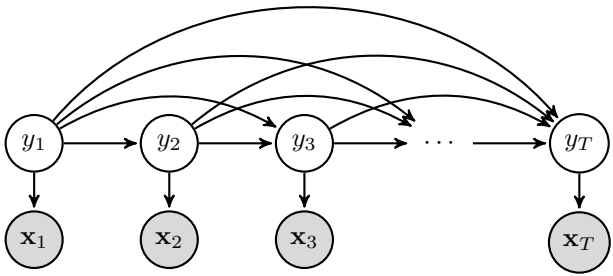
\begin{figure}[h]
\centering
\begin{tikzpicture}[->,>=stealth',shorten >=1pt,auto,node distance=1cm,thick]
  \tikzstyle{every state}=[fill=none,draw=black,text=black,minimum size=0.75cm]
  \tikzstyle{obs}=[fill=black!15,node distance=0.5cm]
  \node[state] (y1)                       {$\y_1$};
  \node[state,right=of y1] (y2)           {$\y_2$};
  \node[state,right=of y2] (y3)           {$\y_3$};
  \node[state,right=of y3,draw=none] (yt) {$\cdots$};
  \node[state,right=of yt] (yT)           {$\y_T$};
  \node[state,obs,below=of y1] (x1)           {$\x_1$};
  \node[state,obs,below=of y2] (x2)           {$\x_2$};
  \node[state,obs,below=of y3] (x3)           {$\x_3$};
  \node[state,obs,below=of yT] (xT)           {$\x_T$};
  \path (y1) edge (x1) edge (y2) edge [bend left] (y3) edge [bend left=40] (yt) edge [bend left=50] (yT);
  \path (y2) edge (x2) edge (y3) edge [bend left] (yt) edge [bend left=40] (yT);
  \path (y3) edge (x3) edge (yt) edge [bend left] (yT);
  \path (yt) edge (yT);
  \path (yT) edge (xT);
\end{tikzpicture}
\caption{Generative chord sequence model. Each chord label $\y_t$ depends
         on all previous labels $\seq{\y}{1}{t-1}$.}
\label{fig:generative_model}
\end{figure}

Chord recognition is a sequence labelling task, i.e.\ we need to assign a
categorical label $\y_t \in \Yset$ (a chord from a chord alphabet) to each
member of the observed sequence $\x_t$ (an audio frame), such that $\y_t$ is
the harmonic interpretation of the music represented by $\x_t$. Formally,
\begin{align}
\seq{\hat{\y}}{1}{T} = \argmax_{\seq{\y}{1}{T}} P\left(\seq{\y}{1}{T} \mid
\seq{\x}{1}{T}\right).\label{eq:decoding_eq}
\end{align}
Assuming a generative structure as shown in
Fig.~\ref{fig:generative_model}, the probability distribution factorises as
$$P(\Y \mid \X) \propto \prod_t \frac{1}{P(\y_t)} P_A\left(\y_t \mid
\x_t\right) P_T\left(\y_t \mid \seq{\y}{1}{t-1}\right),$$ where $P_A$ is the
acoustic model, $P_T$ the temporal model, and $P(y_t)$ the label prior which
we assume to be uniform as in~\cite{renals_connectionist_1994}.

The temporal model $P_T$ predicts the chord symbol of each audio frame. As
discussed earlier, this prevents both finite-context models (such as HMMs or
CRFs) and unrestricted models (such as RNNs) to learn meaningful harmonic
relations. To enable this, we disentangle $P_T$ into a \emph{harmonic language
model} $P_L$ and a \emph{duration model} $P_D$, where the former models the
harmonic progression of a piece, and the latter models the duration of chords.

The language model $P_L$ is defined as $P_L\left(\bar{\y}_k \mid \seq{\bar{\y}}{1}{k-1}\right)$,
where $\seq{\bar{\y}}{1}{k} = \compress{\seq{\y}{1}{t}}$, and $\compress{\cdot}$ is a
sequence compression mapping that removes all consecutive duplicates of a chord
(e.g.\, $\compress{(C,C,F,F,G)} = (C,F,G)$). The frame-wise labels
$\seq{y}{1}{t}$ are thus reduced to chord changes, and $P_L$ can focus on
modelling these.

The duration model $P_D$ is defined as $P_D\left(\s_t \mid
\seq{\y}{1}{t-1}\right)$, where $\s_t \in \{\text{c}, \text{s}\}$ indicates
whether the chord changes (c) or stays the same (s) at time $t$. $P_D$ thus
only predicts whether the chord will change or not, but not which chord will
follow---this is left to the language model $P_L$. This definition allows
$P_D$ to consider the preceding \emph{chord labels} $\seq{y}{1}{t-1}$; in
practice, we restrict the model to only depend on the preceding chord \emph{changes},
i.e.\ $P_D\left(\s_t \mid \seq{s}{1}{t-1}\right)$. Exploring more
complex models of harmonic rhythm is left for future work.

Using these definitions, the temporal model $P_T$ factorises as
\newcommand{\past}{\seq{\y}{1}{t-1}}
\newcommand{\pastchords}{\seq{\bar{\y}}{1}{k-1}}
\begin{align}
P_T&\left(\y_t \mid \past\right) = \label{eq:temporal_model}\\
	&\begin{cases}
P_L\left(\bar{\y}_k \mid \pastchords\right) P_D\left(\text{c} \mid \past \right) & \text{if } y_t \neq y_{t-1} \\
     P_D\left(\text{s} \mid \past \right) & \text{else}
     \end{cases}.\nonumber
\end{align}
The chord progression can then be interpreted as a path through a chord-time
lattice as shown in Fig.~\ref{fig:chord_time_lattice}.

This model cannot be decoded efficiently at test-time because each $y_t$ depends
on all predecessors. We will thus use either models that restrict these
connections to a finite past (such as higher-order Markov models) or use
approximate inference methods for other models (such as RNNs).

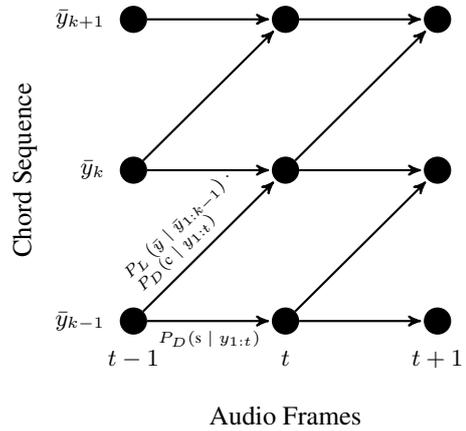
\begin{figure}[t]
\centering
\begin{tikzpicture}[->,>=stealth',shorten >=1pt,auto,node distance=2cm,thick]
\tikzstyle{every state}=[fill=black,draw=black,text=black,minimum size=0.25cm]
\tikzstyle{axis}=[fill=none,draw=none,text=black,font=\small,node distance=0.75cm,minimum width=1cm]
\node[state] (00)                    {};
\node[state] (01) [right of=00]      {};
\node[state] (02) [right of=01]      {};
\node[state] (10) [above of=00]      {};
\node[state] (11) [right of=10]      {};
\node[state] (12) [right of=11]      {};
\node[state] (20) [above of=10]      {};
\node[state] (21) [right of=20]      {};
\node[state] (22) [right of=21]      {};
\path (00) edge node[font=\tiny,sloped,below] {$P_D(\text{s}\mid\seq{\y}{1}{t})$} (01)
             edge node[font=\tiny,sloped,pos=0.9,align=left] {$P_L\left(\bar{\y} \mid \pastchords\right) \cdot$ \\ $P_D(\text{c}\mid\seq{\y}{1}{t})$} (11)
        (01) edge node {} (02)
             edge node {} (12)
        (10) edge node {} (11)
             edge node {} (21)
        (11) edge node {} (12)
             edge node {} (22)
        (20) edge node {} (21)
        (21) edge node {} (22);

  \node[axis] (tm1) [below of=00,node distance=0.5cm] {$t-1$};
  \node[axis] (t)   [below of=01,node distance=0.5cm] {$t$};
  \node[axis] (tp1) [below of=02,node distance=0.5cm] {$t+1$};
  \node[axis] (km1) [left of=00,text width=0.75cm,align=right] {$\bar{\y}_{k-1}$};
  \node[axis] (k)   [left of=10,text width=0.75cm,align=right] {$\bar{\y}_k$};
  \node[axis] (kp1) [left of=20,text width=0.75cm,align=right] {$\bar{\y}_{k+1}$};
  \node[fill=none,draw=none,node distance=1.25cm] (AF) [below of=01] {Audio Frames};
  \node[fill=none,draw=none,node distance=1.25cm,left=of 10,rotate=90,anchor=center] (AF) {Chord Sequence};
\end{tikzpicture}
\caption{Chord-time lattice representing the temporal model $P_T$, split
         into a language model $P_L$ and duration model $P_D$. Here,
         $\seq{\bar{\y}}{1}{K}$ represents a concrete
         chord sequence. For each audio frame, we move along the time-axis
         to the right. If the chord changes, we move diagonally to the upper
         right. This corresponds to the first case in
         Eq.~\ref{eq:temporal_model}. If the chord stays the same, we move only
         to the right. This corresponds to the second case of the equation.}
\label{fig:chord_time_lattice}
\end{figure}

\section{Models}

The general model described above requires three sub-models: an acoustic model
$P_A$ that predicts a chord distribution from each audio frame, a duration model
$P_D$ that predicts when chords change, and a language model $P_L$ that predicts
the progression of chords in the piece.

\subsection{Acoustic Model}

The acoustic model we use is a VGG-style convolutional neural network, similar
to the one presented in~\cite{korzeniowski_fully_2016}. It uses three
convolutional blocks: the first consists of 4 layers of 32 3$\times$3 filters
(with zero-padding in each layer), followed by $2\times1$ max-pooling in
frequency; the second comprises 2 layers of 64 such filters followed by the
same pooling scheme; the third is a single layer of 128 12$\times$9 filters.
Each of the blocks is followed by feature-map-wise dropout with probability
0.2, and each layer is followed by batch normalisation~\cite{ioffe_batch_2015}
and an ELU activation function~\cite{clevert_fast_2016}. Finally, a linear
convolution with 25 1$\times$1 filters followed by global average pooling and
a softmax produces the chord class probabilities $P_A(\y_t \mid \x_t)$.


The input to the network is a
\SI[round-mode=off]{1.5}{\second} patch of a quarter-tone spectrogram computed using a
logarithmically spaced triangular filter bank. Concretely, we process the audio
at a sample rate of \SI{44100}{\Hz} using the STFT with a frame size of 8192
and a hop size of 4410. Then, we apply to the magnitude of the STFT a
triangular filter bank with 24 filters per octave between \SI{65}{\Hz} and
\SI{2100}{\Hz}. Finally, we take the logarithm of the resulting magnitudes to
compress the input range.

Neural networks tend to produce over-confident predictions, which in further
consequence could over-rule the predictions of a temporal
model~\cite{chorowski_better_2016}. To mitigate this, we use two techniques:
first, we train the model using uniform smoothing (i.e.\, we assign a proportion
of $1-\beta$ to other classes during training); second, during inference, we apply the
\emph{temperature softmax} function
$ {\sigma_\tau\left(\mathbf{z}\right)}_j =
  \nicefrac{e^{\nicefrac{z_j}{\tau}}}{\sum_{k=1}^K e^{\nicefrac{z_k}{\tau}}}$
instead of the standard softmax in the final layer. Higher values of $\tau$ produce smoother
probability distributions. In this paper, we use $\beta = 0.9$ and $\tau = 1.3$,
as determined in preliminary experiments.

\subsection{Language Model}

The language model $P_L$ predicts the next chord, regardless of its
duration, given the chord sequence it has previously seen. As shown
in~\cite{korzeniowski_largescale_2018}, RNN-based models perform better than
n-gram models at this task. We thus adopt this approach,
and refer the reader to~\cite{korzeniowski_largescale_2018} for details.

To give an overview, we follow the set-up introduced by~\cite{mikolov_recurrent_2010}
and use a recurrent neural network for next-chord prediction. The network's
task is to compute a probability distribution over all possible next
chord symbols, given the chord symbols it has observed before.
Figure~\ref{fig:rnn_next_step_prediction} shows an RNN in a general next-step
prediction task. In our case, the inputs $z_k$ are the chord symbols given by
$\compress{\seq{y}{1}{T}}$.

\newcommand{\hidst}[1]{$\mathbf{h}_{#1}$}
\newcommand{\vecin}[1]{$\textbf{v}(z_{#1})$}
\newcommand{\probout}[1]{$P(z_{#1} \mid \mathbf{h}_{#1})$}

\begin{figure}[t!]
\centering
\begin{tikzpicture}[->,>=stealth',shorten >=1pt,auto,thick]
\tikzstyle{every state}=[fill=none,draw=black,text=black,minimum size=0.9cm,
						 font=\small,inner sep=1pt, node distance=1.4cm]
\tikzstyle{timedist}=[node distance=1.9cm]
\node[state]             (h0) {\hidst{0}};
\node[state,right of=h0,timedist] (h1) {\hidst{1}};
\node[state,right of=h1,timedist] (h2) {\hidst{2}};
\node[state,right of=h2,draw=none] (ht) {$\cdots$};
\node[state,right of=ht] (hK) {\hidst{K}};

\node[state,below of=h1] (v1) {\vecin{0}};
\node[state,below of=h2] (v2) {\vecin{1}};
\node[state,below of=hK,ellipse] (vK) {\vecin{K-1}};

\node[state,above of=h1,ellipse,inner sep=-5pt] (o1) {\probout{1}};
\node[state,above of=h2,ellipse,inner sep=-5pt] (o2) {\probout{2}};
\node[state,above of=hK,ellipse,inner sep=-5pt] (oK) {\probout{K}};

\path (h0) edge (h1)
      (h1) edge (h2) edge (o1)
      (h2) edge (ht) edge (o2)
      (ht) edge (hK)
      (hK) edge (oK)
      (v1) edge (h1)
      (v2) edge (h2)
      (vK) edge (hK);
\end{tikzpicture}
\caption{Sketch of a RNN used for next step prediction, where $z_k$ refers
         to an arbitrary categorical input, $\mathbf{v}(\cdot)$ is a
         (learnable) input embedding vector, and $\mathbf{h}_k$ the hidden state
         at step $k$. Arrows denote matrix multiplications followed by a
         non-linear activation function. The input is padded with a dummy input
         $z_0$ in the beginning. The network then computes the probability
         distribution for the next symbol.
         }
\label{fig:rnn_next_step_prediction}
\end{figure}
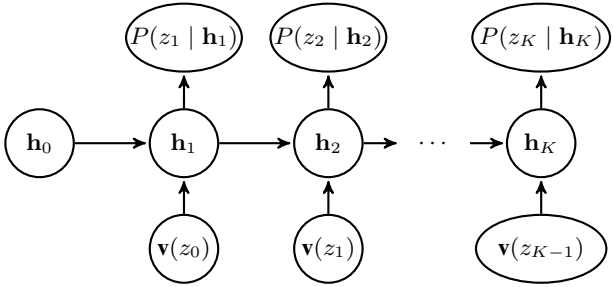

We will describe in detail the network's hyper-parameters in
Section~\ref{sec:experiments}, where we will also evaluate the effect the
language models have on chord recognition.

\subsection{Duration Model}

The duration model $P_D$ predicts whether the chord
will change in the next time step. This corresponds to modelling the duration
of chords. Existing temporal models induce implicit duration models: for
example, an HMM implies an exponential chord duration distribution (if one
state is used to model a chord), or a negative binomial distribution (if
multiple left-to-right states are used per chord). However, such duration
models are simplistic, static, and do not adapt to the processed piece.

An explicit duration model has been explored in~\cite{chen_chord_2012}, where
beat-synchronised chord durations were stored as discrete distributions. Their
approach is useful for beat-synchronised models, but impractical for
frame-wise models---the probability tables would become too large, and data too
sparse to estimate them. Since our approach avoids the potentially error-prone
beat synchronisation, the approach of~\cite{chen_chord_2012} does not work in
our case.

Instead, we opt to use recurrent neural networks to model chord durations.
These models are able to adapt to characteristics of the processed
data~\cite{korzeniowski_largescale_2018}, and have shown great potential in
processing periodic signals~\cite{bock_enhanced_2011} (and chords do change
periodically within a piece). To train an RNN-based duration model, we set up a
next-step-prediction task, identical in principle to the set-up for harmonic
language modelling: the network has to compute the probability of a chord
change in the next time step, given the chord changes it has seen in the past.
We thus simplify $P_D(s_t\mid\seq{\y}{1}{t-1}) \widehat{=}
P_D(s_t\mid\seq{s}{1}{t-1})$, as mentioned earlier. Again, see
Fig.~\ref{fig:rnn_next_step_prediction} for an overview (for duration
modelling, replace $z_k$ with $s_t$).

In Section~\ref{sec:experiments}, we will describe in detail the
hyper-parameters of the networks we employed, and compare the properties of
various settings to baseline duration models. We will also assess the impact on
the duration modelling quality on the final chord recognition result.

\subsection{Model Integration}\label{sec:model_integration}

Dynamic models such as RNNs have one main advantage over their static
counter-parts (e.g.\, n-gram models for language modelling or HMMs for duration
modelling): they consider all previous observations when predicting the next
one. As a consequence, they are able to adapt to the piece that is currently
processed---they assign higher probabilities to sub-sequences of chords
they have seen earlier~\cite{korzeniowski_largescale_2018}, or predict chord
changes according to the harmonic rhythm of a song (see
Sec.~\ref{sec:experiments_duration_models}). The flip side of the coin is,
however, that this property prohibits the use of dynamic programming
approaches for efficient decoding. We cannot exactly \emph{and} efficiently
decode the best chord sequence given the input audio.

Hence we have to resort to approximate inference. In particular, we employ
\emph{hashed beam search}~\cite{sigtia_audio_2015} to decode the chord
sequence. General beam search restricts the search space by keeping only the
$N_b$ best solutions up to the current time step. (In our case, the $N_b$ best
paths through all possible chord-time lattices, see
Fig.~\ref{fig:chord_time_lattice}.) However, as pointed out
in~\cite{sigtia_audio_2015}, the beam might saturate with almost identical
solutions, e.g.\ the same chord sequence differing only marginally in the times
the chords change. Such pathological cases may impair the final estimate. To
mitigate this problem, hashed beam search forces the tracked solutions to be
diverse by pruning similar solutions with lower probability.

The similarity of solutions is determined by a task-specific \emph{hash
function}. For our purpose, we define the hash function of a solution to be
the last $N_h$ chord symbols in the sequence, regardless of their duration;
formally, the hash function $f_h\left(\seq{\y}{1}{t}\right) = \seq{\bar{\y}}{(k-N_h)}{k}$.
(Recall that $\seq{\bar{\y}}{1}{k} = \compress{\seq{\y}{1}{t}}$.)
In contrast to the hash function
originally proposed in~\cite{sigtia_audio_2015}, which directly uses
$\seq{\y}{(t-N_h)}{t}$, our formulation ensures that
sequences that differ only in timing, but not in chord sequence, are considered
similar.

To summarise, we approximately decode the optimal chord transcription as
defined in Eq.~\ref{eq:decoding_eq} using hashed beam search, which at each
time step keeps the best $N_b$ solutions, and at most $N_s$ similar solutions.

\section{Experiments}
\label{sec:experiments}

In our experiments, we will first evaluate harmonic language and duration
models individually. Here, we will compare the proposed models to common
baselines. Then, we will integrate these models into the chord recognition
framework we outlined in Section~\ref{sec:chord_sequence_modelling}, and evaluate
how the individual parts interact in terms of chord recognition score.

\subsection{Data}

We use the following datasets in 4-fold cross-validation.
\textbf{Isophonics\footnote{\url{http://isophonics.net/datasets}}:} 180 songs
by the Beatles, 19 songs by Queen, and 18 songs by Zweieck, 10:21 hours of
audio; \textbf{RWC Popular~\cite{goto_rwc_2002}:} 100 songs in the style of
American and Japanese pop music, 6:46 hours of audio; \textbf{Robbie
Williams~\cite{digiorgi_automatic_2013}:} 65 songs by Robbie Williams, 4:30 of
audio; and \textbf{McGill Billboard~\cite{burgoyne_expert_2011}:} 742 songs
sampled from the American billboard charts between 1958 and 1991, 44:42 hours
of audio. The compound dataset thus comprises 1125 unique songs, and a total
of 66:21 hours of audio.

Furthermore, we used the following data sets (with duplicate songs removed) as
additional data for training the language and duration models:
173 songs from the \textbf{Rock}~\cite{declercq_corpus_2011} corpus; a subset of 160 songs from \textbf{UsPop2002}\footnote{\url{
https://labrosa.ee.columbia.edu/projects/musicsim/uspop2002.html}} for which
chord annotations are available\footnote{\url{https://github.com/tmc323/Chord-
Annotations}}; 291 songs from \textbf{Weimar Jazz}\footnote{\url{http://jazzomat.hfm-
weimar.de/dbformat/dboverview.html}}, with chord annotations taken from lead
sheets of Jazz standards; and \textbf{Jay Chou}~\cite{deng_automatic_2016}, a
small collection of 29 Chinese pop songs. 

We focus on the major/minor chord vocabulary, and
following~\cite{cho_relative_2014}, map all chords containing a minor third to
minor, and all others to major. This leaves us with 25 classes: $\text{12 root
notes} \times \{\text{major}, \text{minor}\}$ and the `no- chord' class.

\subsection{Language Models}

The performance of neural networks depends on a good choice of hyper-parameters,
such as number of layers, number of units per layer, or unit type (e.g.\,
vanilla RNN, gated recurrent unit (GRU)~\cite{cho_properties_2014} or long
short-term memory unit (LSTM)~\cite{hochreiter_long_1997}). The findings
in~\cite{korzeniowski_largescale_2018} provide a good starting point for
choosing hyper-parameter settings that work well. However, we strive to find a
simpler model to reduce the computational burden at test time. To this end, we
perform a grid search in a restricted search space, using the validation score
of the first fold. We search over the following settings: number of layers
$\in \{1, 2, 3\}$, number of units $\in\{256, 512\}$, unit type
$\in\{\text{GRU}, \text{LSTM}\}$, input embedding $\in\{\text{one-hot},
\mathbb{R}^8, \mathbb{R}^{16}, \mathbb{R}^{24}\}$, learning rate $\in \{0.001,
0.005\}$, and skip connections $\in \{ \text{on}, \text{off}\}$. Other
hyper-parameters were fixed for all trials: we train the networks for 100
epochs using stochastic gradient descent with mini-batches of size 4, employ
the Adam update rule~\cite{kingma_adam_2015}, and starting from epoch 50,
linearly anneal the learning rate to 0.

To increase the diversity in the training data, we use two data augmentation
techniques, applied each time we show a piece to the network. First, we
randomly shift the key of the piece; the network can thus learn that harmonic
relations are independent of the key, as in roman numeral analysis. Second, we
select a sub-sequence of random length instead of the complete chord sequence;
the network thus has to learn to cope with varying context sizes.

\begin{table}[t]
\centering
\begin{tabular}{@{}lrrrr@{}}
\toprule
      & \textit{GRU-512}     & \textit{GRU-32}      & \textit{4-gram}      & \textit{2-gram}       \\ \midrule
log-P & \num{-1.29345285892} & \num{-1.57605493069} & \num{-1.88675260544} & \num{-2.39250016212}  \\ \bottomrule
\end{tabular}
\caption{Language model results: average log-probability of the correct next chord computed by each model.}
\label{tab:lang_mod_results}
\end{table}

The best model turned out to be a single-layer network of 512 GRUs, with a
learnable 16-dimensional input embedding and without skip connections, trained
using a learning rate of 0.005\footnote{Due to space constraints, we cannot
present the complete grid search results.}. We compare this model and a
smaller, but otherwise identical RNN with 32 units, to two baselines: a 2-gram
model, and a 4-gram model. Both can be used for chord recognition in a
higher-order HMM~\cite{korzeniowski_automatic_2018}. We train the n-gram
models using maximum likelihood estimation with Lidstone smoothing as described
in~\cite{korzeniowski_largescale_2018}, using the key-shift data augmentation
technique (sub-sequence cropping is futile for finite context models). As
evaluation measure, we use the average log-probability of predicting the
correct next chord. Table~\ref{tab:lang_mod_results} presents the test results.
The GRU models predict chord sequences with much higher probability than the
baselines.

\begin{figure}[t!]
	\centering
	\begin{minipage}[c]{0.5\columnwidth}
    \input{figs/chord_embedding.pgf}
    \end{minipage}
    \hfill
    \begin{minipage}[c]{0.4\columnwidth}
	\includegraphics[width=\textwidth]{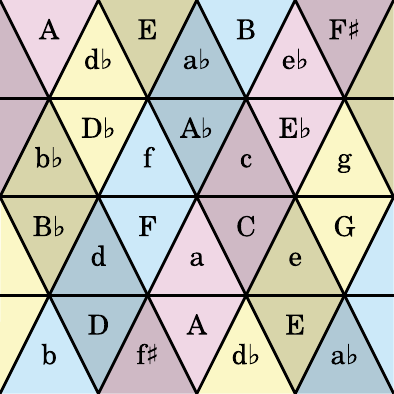}
    \end{minipage}
	\caption{
	Chord embedding projected into 2D using PCA (left); Tonnetz of triads
	(right). The ``no-chord'' class resides in the center of the embedding.
	Major chords are upper-case and orange, minor chords lower-case and blue.
	Clusters in the projected embedding and the corresponding positions in the
	Tonnetz are marked in color. If projected into 3D (not shown here), the
	chord clusters split into a lower and upper half of four chords each. The
	chords in the lower halves are shaded in the Tonnetz representation.
	}
	\label{fig:chord_embedding}
\end{figure}

When we look into the input embedding $\mathbf{v}(\cdot)$, which was learned
by the RNN during training from a random initialisation, we observe an
interesting positioning of the chord symbols (see
Figure~\ref{fig:chord_embedding}). We found
that similar patterns develop for all 1-layer GRUs we tried, and these
patterns are consistent for all folds we trained on.
We observe
\begin{enumerate*}[label=\roman*)]
\item that chords form three clusters around the center, in which the minor chords are farther from the	center than major chords;
\item that the clusters group major and minor
	chords with the same root, and the distance between the roots are minor
	thirds (e.g.\, C, E$\flat$, F$\sharp$, A);
\item that clockwise movement
	in the circle of fifths corresponds to clockwise movement in the projected
	embedding; and
\item that the way chords are grouped in the embedding corresponds to
      how they are connected in the Tonnetz.
\end{enumerate*}

At this time, we cannot
provide an explanation for these automatically emerging patterns. However, they
warrant a further investigation to uncover why this specific arrangement seems
to benefit the predictions of the model.

\subsection{Duration Models}
\label{sec:experiments_duration_models}

As for the language model, we performed a grid search on the first fold to
find good choices for the recurrent unit type $\in\{\text{vanilla RNN},
\text{GRU}, \text{LSTM}\}$, and number of recurrent units $\in \{16, 32, 64,
128, 256\}$ for the LSTM and GRU, and $\{128, 256, 512\}$ for the vanilla RNN.
We use only one recurrent layer for simplicity. We found networks of 256 GRU
units to perform best; although this indicates that even bigger models might
give better results, for the purposes of this study, we think that this
configuration is a good balance between prediction quality and model
complexity.

The models were trained for 100 epochs using the Adam update
rule~\cite{kingma_adam_2015} with a learning rate linearly decreasing from
0.001 to 0. The data was processed in mini-batches of 10, where the sequences
were cut in excerpts of 200 time steps (\SI{20}{\second}). We also applied
gradient clipping at a value of 0.001 to ensure a smooth learning progress.

We compare the best RNN-based duration model with two baselines. The baselines
are selected because both are implicit consequences of using HMMs as temporal
model, as it is common in chord recognition. We assume a single
parametrisation for each chord; this ostensible simplification is justified,
because simple temporal models such as HMMs do not profit from chord
information, as shown by~\cite{cho_relative_2014,chen_chord_2012}. The first
baseline we consider is a \emph{negative binomial distribution}. It can be
modelled by a HMM using $n$ states per chord, connected in a left-to-right
manner, with transitions of probability $p$ between the states (self-transitions
thus have probability $1-p$). The second, a special case of the
first with $n=1$, is an \emph{exponential distribution}; this is the implicit
duration distribution used by all chord recognition models that employ a
simple 1-state-per-chord HMM as temporal model. Both baselines are trained
using maximum likelihood estimation.

To measure the quality of a duration model, we consider the average
log-probability it assigns to a chord duration. The results are shown in
Table~\ref{tab:dur_mod_results}. We further added results for the simplest GRU
model we tried---using only 16 recurrent units---to indicate the performance of
small models of this type. We will also use this simple model when judging the
effect of duration modelling on the final result in
Sec.~\ref{sec:integrated_models}. As seen in the table, both GRU models clearly
out-perform the baselines.

\begin{figure}[t!]
\centering
\input{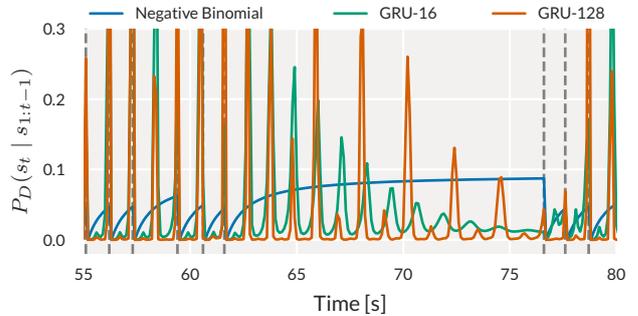}
\vspace{-2em}
\caption{Probability of chord change computed by different models.
         Gray vertical dashed lines indicate true chord changes.}
\label{fig:duration_predictions}
\end{figure}

\begin{table}[t]
\centering
\begin{tabular}{@{}lrrrr@{}}
\toprule

      & \textit{GRU-256} & \textit{GRU-16} & \textit{Neg. Binom.} & \textit{Exp.} \\ \midrule
log-P & \num{-2.013729}  & \num{-2.867829} & \num{-3.946390}            & \num{-4.003020}      \\ \bottomrule
\end{tabular}
\caption{Duration model results: average log-probability of
chord durations computed by each model.}
\label{tab:dur_mod_results}
\end{table}

Figure~\ref{fig:duration_predictions} shows the reason why the GRU performs so
much better than the baselines: as a dynamic model, it can adapt to the harmonic
rhythm of a piece, while static models are not capable of doing so. We see
that a GRU with 128 units predicts chord changes with high probability at
periods of the harmonic rhythm. It also reliably remembers the period over
large gaps in which the chord did not change (between seconds 61 and 76).
During this time, the peaks decay differently for different multiples of the
period, which indicates that the network simultaneously tracks multiple
periods of varying importance. In contrast, the negative binomial distribution
statically yields a higher chord change probability that rises with the
number of audio frames since the last chord change. Finally, the smaller GRU
model with only 16 units also manages to adapt to the harmonic rhythm;
however, its predictions between the peaks are noisier, and it
fails to remember the period correctly in the time without chord changes.

\subsection{Integrated Models}
\label{sec:integrated_models}

The individual results for the language and duration models are encouraging,
but only meaningful if they translate to better chord recognition scores. This
section will thus evaluate if and how the duration and language models affect
the performance of a chord recognition system.

The acoustic model used in these experiments was trained for 300 epochs (with
200 parameter updates per epoch) using a mini-batch size of 512 and the Adam
update rule with standard parameters. We linearly decay the learning rate to
0 in the last 100 epochs.

We compare all combinations of language and duration models presented in the
previous sections. For language modelling, these are the GRU-512, GRU-32, 4-gram,
and 2-gram models; for duration modelling, these are the GRU-256, GRU-16,
and negative binomial models. (We leave out the exponential model, because its
results differ negligibly from the negative binomial one).
The models are decoded using the Hashed Beam
Search algorithm, as described in Sec.~\ref{sec:model_integration}: we
use a beam width of $N_b = 25$, where we track at most $N_s = 4$ similar solutions
as defined by the hash function $f_h$, where the number of chords considered is
set to $N_h = 5$. These values were determined by a small number of preliminary
experiments.

Additionally, we evaluate exact decoding results for the n-gram language
models in combination with the negative binomial duration
distribution. This will indicate how much the results suffer due to the
approximate beam search.

As main evaluation metric, we use the weighted chord symbol recall (WCSR) over the major/minor chord alphabet, as defined in~\cite{pauwels_evaluating_2013}.
We thus compute $\text{WCSR} = \nicefrac{t_c}{t_a}$, where $t_c$ is the total
duration of chord segments that have been recognised correctly, and $t_a$ is
the total duration of chord segments annotated with chords from the target
alphabet. We also report chord root accuracy and a measure of segmentation
(see~\cite{harte_automatic_2010}, Sec. 8.3). Table~\ref{tab:results} compares
the results of the standard model (the combination that implicitly emerges in
simple HMM-based temporal models) to the best model found in this study.
Although the improvements are modest, they are consistent, as shown by
a paired t-test ($p < \num{2.48680337047e-23}$ for all differences).

\begin{table}[t]
\centering
\begin{tabular}{@{}lrrr@{}}
\toprule
\textit{Model}                     & \textit{Root} & \textit{Maj/Min} & \textit{Seg.} \\ \midrule
2-gram / neg. binom. & \num{0.812297}     & \num{0.795497}          & \num{0.80408}         \\
GRU-512 / GRU-256       & \textbf{\num{0.821112}}     & \textbf{\num{0.80473}}           & \textbf{\num{0.81392}}         \\ \bottomrule
\end{tabular}
\caption{Results of the standard model (2-gram language model with negative binomial durations) compared to the best one (GRU language and duration models).}
\label{tab:results}
\end{table}

\begin{figure}[t!]
\centering
\input{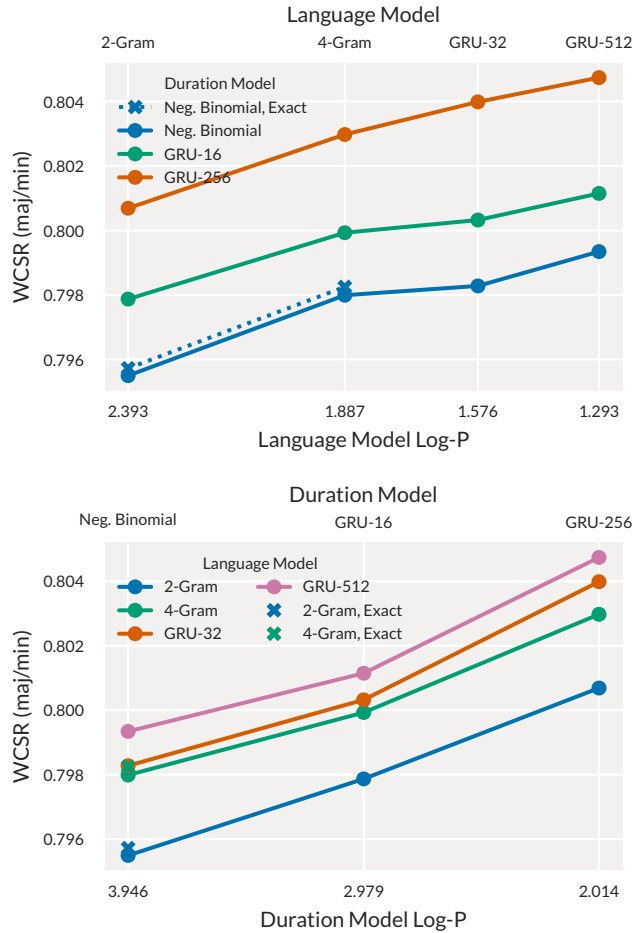}
\vspace{-2em}
\caption{Effect of language and duration models on the final result. Both
plots show the same results from different perspectives.}
\label{fig:final_results}
\end{figure}

Figure~\ref{fig:final_results} presents the effects of duration and language
models on the WCSR.\@ Better language and duration models directly improve
chord recognition results, as the WCSR increases linearly with higher
log-probability of each model.  As this relationship does not seem to flatten
out, further improvement of each model type can still increase the score. We
also observe that the approximate beam search does not impair the result
by much compared to exact decoding (compare the dotted blue line 
with the solid one).

\section{Conclusion and Discussion}

We described a probabilistic model that disentangles three components of a
chord recognition system: the acoustic model, the duration model, and the
language model. We then developed better duration and language models than have
been used for chord recognition, and illustrated why the RNN-based duration
models perform better and are more meaningful than their static counterparts
implicitly employed in HMMs. (For a similar investigation for chord language
models, see~\cite{korzeniowski_largescale_2018}.) Finally, we showed that
improvements in each of these models directly influence chord recognition
results.

We hope that our contribution facilitates further research in harmonic language
and duration models for chord recognition. These aspects have been neglected
because they did not show great potential for improving the final
result~\cite{cho_relative_2014,chen_chord_2012}. However, we believe
(see~\cite{korzeniowski_futility_2017} for some evidence) that this was due to
the improper assumption that temporal models applied on the time-frame level
can appropriately model musical knowledge. The results in this paper
indicate that chord transitions modelled on the chord level, and connected to
audio frames via strong duration models, indeed have the capability to improve
chord recognition results.

\section{Acknowledgements}

This work is supported by the European Research Council (ERC) under the EU's
Horizon 2020 Framework Programme (ERC Grant Agreement number 670035, project
``Con Espressione'').

\bibliography{ismir2018}

\begin{thebibliography}{10}

\bibitem{bock_enhanced_2011}
Sebastian B{\"o}ck and Markus Schedl.
\newblock Enhanced {{Beat Tracking With Context}}-{{Aware Neural Networks}}.
\newblock In {\em Proceedings of the 14th {{International Conference}} on
  {{Digital Audio Effects}} ({{DAFx}}-11)}, Paris, France, September 2011.

\bibitem{boulanger-lewandowski_audio_2013}
Nicolas Boulanger-Lewandowski, Yoshua Bengio, and Pascal Vincent.
\newblock Audio chord recognition with recurrent neural networks.
\newblock In {\em Proceedings of the 14th {{International Society}} for {{Music
  Information Retrieval Conference}} ({{ISMIR}})}, Curitiba, Brazil, 2013.

\bibitem{burgoyne_expert_2011}
John~Ashley Burgoyne, Jonathan Wild, and Ichiro Fujinaga.
\newblock An {{Expert Ground Truth Set}} for {{Audio Chord Recognition}} and
  {{Music Analysis}}.
\newblock In {\em Proceedings of the 12th {{International Society}} for {{Music
  Information Retrieval Conference}} ({{ISMIR}})}, Miami, USA, October 2011.

\bibitem{chen_chord_2012}
Ruofeng Chen, Weibin Shen, Ajay Srinivasamurthy, and Parag Chordia.
\newblock Chord {{Recognition Using Duration}}-{{Explicit Hidden Markov
  Models}}.
\newblock In {\em Proceedings of the 13th {{International Society}} for {{Music
  Information Retrieval Conference}} ({{ISMIR}})}, Porto, Portugal, 2012.

\bibitem{cho_properties_2014}
Kyunghyun Cho, Bart {van Merrienboer}, Dzmitry Bahdanau, and Yoshua Bengio.
\newblock On the {{Properties}} of {{Neural Machine Translation}}:
  {{Encoder}}-{{Decoder Approaches}}.
\newblock {\em arXiv:1409.1259 [cs, stat]}, September 2014.

\bibitem{cho_improved_2014}
Taemin Cho.
\newblock {\em Improved {{Techniques}} for {{Automatic Chord Recognition}} from
  {{Music Audio Signals}}}.
\newblock Dissertation, New York University, New York, 2014.

\bibitem{cho_relative_2014}
Taemin Cho and Juan~P. Bello.
\newblock On the {{Relative Importance}} of {{Individual Components}} of
  {{Chord Recognition Systems}}.
\newblock {\em IEEE/ACM Transactions on Audio, Speech, and Language
  Processing}, 22(2):477--492, February 2014.

\bibitem{cho_exploring_2010}
Taemin Cho, Ron~J Weiss, and Juan~Pablo Bello.
\newblock Exploring common variations in state of the art chord recognition
  systems.
\newblock In {\em Proceedings of the {{Sound}} and {{Music Computing
  Conference}} ({{SMC}})}, pages 1--8, 2010.

\bibitem{chorowski_better_2016}
Jan Chorowski and Navdeep Jaitly.
\newblock Towards better decoding and language model integration in sequence to
  sequence models.
\newblock {\em arXiv:1612.02695 [cs, stat]}, December 2016.

\bibitem{clevert_fast_2016}
Djork-Arn{\'e} Clevert, Thomas Unterthiner, and Sepp Hochreiter.
\newblock Fast and {{Accurate Deep Network Learning}} by {{Exponential Linear
  Units}} ({{ELUs}}).
\newblock In {\em International {{Conference}} on {{Learning Representations}}
  ({{ICLR}}), {{arXiv}}:1511.07289}, San Juan, Puerto Rico, February 2016.

\bibitem{declercq_corpus_2011}
Trevor {de Clercq} and David Temperley.
\newblock A corpus analysis of rock harmony.
\newblock {\em Popular Music}, 30(01):47--70, January 2011.

\bibitem{deng_automatic_2016}
Junqi Deng and Yu-Kwong Kwok.
\newblock Automatic {{Chord}} estimation on seventhsbass {{Chord}} vocabulary
  using deep neural network.
\newblock In {\em International {{Conference}} on {{Acoustics Speech}} and
  {{Signal Processing}} ({{ICASSP}})}, Shanghai, China, March 2016.

\bibitem{digiorgi_automatic_2013}
Bruno Di~Giorgi, Massimiliano Zanoni, Augusto Sarti, and Stefano Tubaro.
\newblock Automatic chord recognition based on the probabilistic modeling of
  diatonic modal harmony.
\newblock In {\em Proceedings of the 8th {{International Workshop}} on
  {{Multidimensional Systems}}}, Erlangen, Germany, 2013.

\bibitem{fujishima_realtime_1999}
Takuya Fujishima.
\newblock Realtime {{Chord Recognition}} of {{Musical Sound}}: A {{System Using
  Common Lisp Music}}.
\newblock In {\em Proceedings of the {{International Computer Music
  Conference}} ({{ICMC}})}, Beijing, China, 1999.

\bibitem{goto_rwc_2002}
Masataka Goto, Hiroki Hashiguchi, Takuichi Nishimura, and Ryuichi Oka.
\newblock {{RWC Music Database}}: {{Popular}}, {{Classical}} and {{Jazz Music
  Databases}}.
\newblock In {\em Proceedings of the 3rd {{International Conference}} on
  {{Music Information Retrieval}} ({{ISMIR}})}, Paris, France, 2002.

\bibitem{harte_automatic_2010}
Christopher Harte.
\newblock {\em Towards {{Automatic Extraction}} of {{Harmony Information}} from
  {{Music Signals}}}.
\newblock Dissertation, Department of Electronic Engineering, Queen Mary,
  University of London, London, United Kingdom, 2010.

\bibitem{hochreiter_long_1997}
Sepp Hochreiter and J{\"u}rgen Schmidhuber.
\newblock Long {{Short}}-{{Term Memory}}.
\newblock {\em Neural Computation}, 9(8):1735--1780, November 1997.

\bibitem{humphrey_rethinking_2012}
Eric~J. Humphrey and Juan~P. Bello.
\newblock Rethinking {{Automatic Chord Recognition}} with {{Convolutional
  Neural Networks}}.
\newblock In {\em 11th {{International Conference}} on {{Machine Learning}} and
  {{Applications}} ({{ICMLA}})}, Boca Raton, USA, December 2012. {IEEE}.

\bibitem{ioffe_batch_2015}
Sergey Ioffe and Christian Szegedy.
\newblock Batch {{Normalization}}: {{Accelerating}} deep network training by
  reducing internal covariate shift.
\newblock {\em arXiv preprint arXiv:1502.03167}, 2015.

\bibitem{kingma_adam_2014}
Diederik Kingma and Jimmy Ba.
\newblock Adam: {{A}} method for stochastic optimization.
\newblock {\em arXiv preprint arXiv:1412.6980}, 2014.

\bibitem{korzeniowski_largescale_2018}
Filip Korzeniowski, David R.~W. Sears, and Gerhard Widmer.
\newblock A {{Large}}-{{Scale Study}} of {{Language Models}} for {{Chord
  Prediction}}.
\newblock In {\em {{IEEE International Conference}} on {{Acoustics}},
  {{Speech}} and {{Signal Processing}} ({{ICASSP}})}, Calgary, Canada, April
  2018.

\bibitem{korzeniowski_feature_2016}
Filip Korzeniowski and Gerhard Widmer.
\newblock Feature {{Learning}} for {{Chord Recognition}}: {{The Deep Chroma
  Extractor}}.
\newblock In {\em Proceedings of the 17th {{International Society}} for {{Music
  Information Retrieval Conference}} ({{ISMIR}})}, New York, USA, August 2016.

\bibitem{korzeniowski_fully_2016}
Filip Korzeniowski and Gerhard Widmer.
\newblock A {{Fully Convolutional Deep Auditory Model}} for {{Musical Chord
  Recognition}}.
\newblock In {\em Proceedings of the {{IEEE International Workshop}} on
  {{Machine Learning}} for {{Signal Processing}} ({{MLSP}})}, Salerno, Italy,
  September 2016.

\bibitem{korzeniowski_futility_2017}
Filip Korzeniowski and Gerhard Widmer.
\newblock On the {{Futility}} of {{Learning Complex Frame}}-{{Level Language
  Models}} for {{Chord Recognition}}.
\newblock In {\em Proceedings of the {{AES International Conference}} on
  {{Semantic Audio}}}, Erlangen, Germany, June 2017.

\bibitem{korzeniowski_automatic_2018}
Filip Korzeniowski and Gerhard Widmer.
\newblock Automatic {{Chord Recognition}} with {{Higher}}-{{Order Harmonic
  Language Modelling}}.
\newblock In {\em Proceedings of the 26th {{European Signal Processing
  Conference}} ({{EUSIPCO}})}, Rome, Italy, September 2018.

\bibitem{mauch_simultaneous_2010}
M.~Mauch and S.~Dixon.
\newblock Simultaneous {{Estimation}} of {{Chords}} and {{Musical Context From
  Audio}}.
\newblock {\em IEEE Transactions on Audio, Speech, and Language Processing},
  18(6):1280--1289, August 2010.

\bibitem{mcfee_structured_2017}
Brian McFee and Juan~Pablo Bello.
\newblock Structured {{Training}} for {{Large}}-{{Vocabulary Chord
  Recognition}}.
\newblock In {\em Proceedings of the 18th {{International Society}} for {{Music
  Information Retrieval Conference}} ({{ISMIR}})}, Suzhou, China, October 2017.

\bibitem{mikolov_recurrent_2010}
Tomas Mikolov, Martin Karafi{\'a}t, Luk{\'a}s Burget, Jan Cernock{\'y}, and
  Sanjeev Khudanpur.
\newblock Recurrent neural network based language model.
\newblock In {\em {{INTERSPEECH}} 2010, 11th {{Annual Conference}} of the
  {{International Speech Communication Association}}, {{Makuhari}}, {{Chiba}},
  {{Japan}}, {{September}} 26-30, 2010}, pages 1045--1048, Chiba, Japan, 2010.

\bibitem{muller_making_2009}
Meinard M{\"u}ller, Sebastian Ewert, and Sebastian Kreuzer.
\newblock Making chroma features more robust to timbre changes.
\newblock In {\em International {{Conference}} on {{Acoustics}}, {{Speech}} and
  {{Signal Processing}} ({{ICASSP}})}. {IEEE}, 2009.

\bibitem{pauwels_evaluating_2013}
Johan Pauwels and Geoffroy Peeters.
\newblock Evaluating automatically estimated chord sequences.
\newblock In {\em 2013 {{IEEE International Conference}} on {{Acoustics}},
  {{Speech}} and {{Signal Processing}}}, pages 749--753. {IEEE}, 2013.

\bibitem{renals_connectionist_1994}
S.~Renals, N.~Morgan, H.~Bourlard, M.~Cohen, and H.~Franco.
\newblock Connectionist {{Probability Estimators}} in {{HMM Speech
  Recognition}}.
\newblock {\em IEEE Transactions on Speech and Audio Processing},
  2(1):161--174, January 1994.

\bibitem{sigtia_audio_2015}
Siddharth Sigtia, Nicolas Boulanger-Lewandowski, and Simon Dixon.
\newblock Audio chord recognition with a hybrid recurrent neural network.
\newblock In {\em 16th {{International Society}} for {{Music Information
  Retrieval Conference}} ({{ISMIR}})}, M{\'a}laga, Spain, October 2015.

\bibitem{ueda_hmmbased_2010}
Yushi Ueda, Yuki Uchiyama, Takuya Nishimoto, Nobutaka Ono, and Shigeki
  Sagayama.
\newblock {{HMM}}-based approach for automatic chord detection using refined
  acoustic features.
\newblock In {\em International {{Conference}} on {{Acoustics Speech}} and
  {{Signal Processing}} ({{ICASSP}})}, Dallas, USA, March 2010.

\end{thebibliography}

\end{document}